\documentclass{tMOP2e}
\usepackage{epsfig}

\begin{document}
\title{Cooperative Scattering by Cold Atoms}

\author{S. Bux$^{a,b}$, E. Lucioni$^{a,c}$, H. Bender$^{b}$, T. Bienaim\'e$^{a}$, K. Lauber$^{b}$,  C. Stehle$^{b}$, C. Zimmermann$^{b}$, S. Slama$^{b}$, Ph.W. Courteille$^{a,b,d}$, N. Piovella$^{c}$ and R. Kaiser$^{a}$$^{\ast}$\vspace{6pt}\thanks{$^\ast$Corresponding author. Email: robin.kaiser@inln.cnrs.fr \vspace{6pt}}
\\\vspace{6pt}
$^{a}${\em{Institut Non Lin\'eaire de Nice, UMR 6618 CNRS, 1361 route des Lucioles, F-06560 Valbonne, France.}};\\
 $^{b}${\em{Physikalisches Institut, Eberhardt-Karls-Universit\"at T\"ubingen, D-72076 T\"ubingen, Germany.}};\\
  $^{c}${\em{Dipartimento di Fisica, Universit\`a Degli Studi di Milano, Via Celoria 16, I-20133 Milano, Italy.}};\\
  $^{d}${\em{Instituto de F\'isica de S\~ao Carlos, Universidade de S\~ao Paulo, 13560-970 S\~ao Carlos, SP, Brazil.}}\\\vspace{6pt}\received{February 2010} }

\maketitle

\begin{abstract}

We have studied the interplay between disorder and cooperative scattering for single scattering limit in the presence of a driving laser. Analytical results have been derived and we have observed cooperative scattering effects in a variety of experiments, ranging from thermal atoms in an optical dipole trap, atoms released from a dark MOT and atoms in a BEC, consistent with our theoretical predictions.

\bigskip
\end{abstract}

\section{Introduction}

The interaction of quasi resonant light with a large cloud of atoms has been studied for many years, starting with seminal work by Dicke \cite{Dicke54} and with renewed interest in the context of entanglement which such systems are expected to contain. Often continuous density distributions have been assumed which allow for analytical expressions to be obtained. The role of fluctuations of the atomic density is however by itself at the origin of interesting phenomena, such as Anderson localisation of light \cite{AGK, Sokolov2009}. In a series of theoretical and experimental studies, we have recently addressed the question of the quasi-resonant interaction of light with clouds of cold atoms, bridging the gap from single atom behavior, effects dominated by disorder to a mean field regime, where a continuous density distribution is the relevant description.

In this paper, we present a theoretical model we use to describe the collective atomic response under continuous excitation of a low intensity light field, with additional details compared to \cite{Courteille09}. We then present experiments which have been performed in complement to those reported in \cite{Bienaime10}, using atoms in a dipole trap as well as atoms in a magnetic trap, both above and below the Bose-Einstein condensation temperature.

\section{Theoretical Description}
\textit{Hamiltonian and  state of the system --} We consider a
cloud of $N$ two level atoms (positions $\mathbf{r}_j$, lower and
upper states $|g_j\rangle$ and $|e_j\rangle$ respectively,
transition frequency $\omega_a$, excited state lifetime
$1/\Gamma$), excited by a quasi-resonant incident laser
propagating along the direction $\mathbf{\hat e}_z$ (wave vector
$\mathbf{k}_0$) and with frequency $\omega_0=\omega_a+\Delta_0$.
The atom-field interaction Hamiltonian is in the rotating-wave
approximation (RWA) \cite{Courteille09, Note01}
\begin{align}\label{Ham}
    \hat{H} =\hbar\sum_{j=1}^N\left[\frac{\Omega_0}{2}\hat{\sigma}_j e^{i\Delta_0t-i\mathbf{k}_0\cdot\mathbf{r}_j}
    +\textrm{h.c.}\right]
    +\hbar\sum_{j=1}^N\sum_{\mathbf{k}}\left[g_k\hat{\sigma}_j\hat{a}_{\mathbf{k}}^\dagger
    e^{i\Delta_kt-i\mathbf{k}\cdot\mathbf{r}_j}
        +\textrm{h.c.}\right].
\end{align}
Here, $\Omega_0$ is the Rabi frequency of the interaction between
an atom and the classical pump mode,
$\hat{\sigma}_j=|g_j\rangle\langle e_j|$ is the lowering operator
for atom $j$, $\hat{a}_{\mathbf{k}}$ is the photon annihilation
operator, and $g_k =d\sqrt{\omega_k/(\hbar\epsilon_0V_{ph})}$
describes the coupling between the atom and the vacuum modes with
volume $V_{ph}$ and frequency $\omega_k=\omega_a+\Delta_k$.  We
assume that all atoms are driven by the unperturbed incident laser
beam, thus neglecting dephasing by atoms along the laser path or
by near field effects, which could arise for large spatial
densities. Calling $|0\rangle_a=|g_1,..,g_N\rangle$ the atomic
ground state and $|j\rangle_a =|g_1,..,e_j,..,g_N\rangle$ the
state where only the atom $j$ is excited, we assume that the total
state of the system has the following form
\cite{Scully06,Svidzinsky08}:
\begin{align}\label{EqAnsatz}
    |\Psi(t)\rangle =\alpha(t)|0\rangle_a|0\rangle_{\mathbf{k}}
    +e^{-i\Delta_0 t}\sum_{j=1}^N\tilde\beta_j(t)
        e^{i \mathbf{k}_0\cdot \mathbf{r}_j}
        |j\rangle_a|0\rangle_{\mathbf{k}}
        +\sum_{\mathbf{k}}\gamma_{\mathbf{k}}(t)|0\rangle_a|1\rangle_{\mathbf{k}}~.
\end{align}
The above expression assumes that only states with at most one
atomic excitation contribute to the effects here described
\cite{Note02}.

\begin{figure}[ht]
        \centerline{\scalebox{.5}{\includegraphics{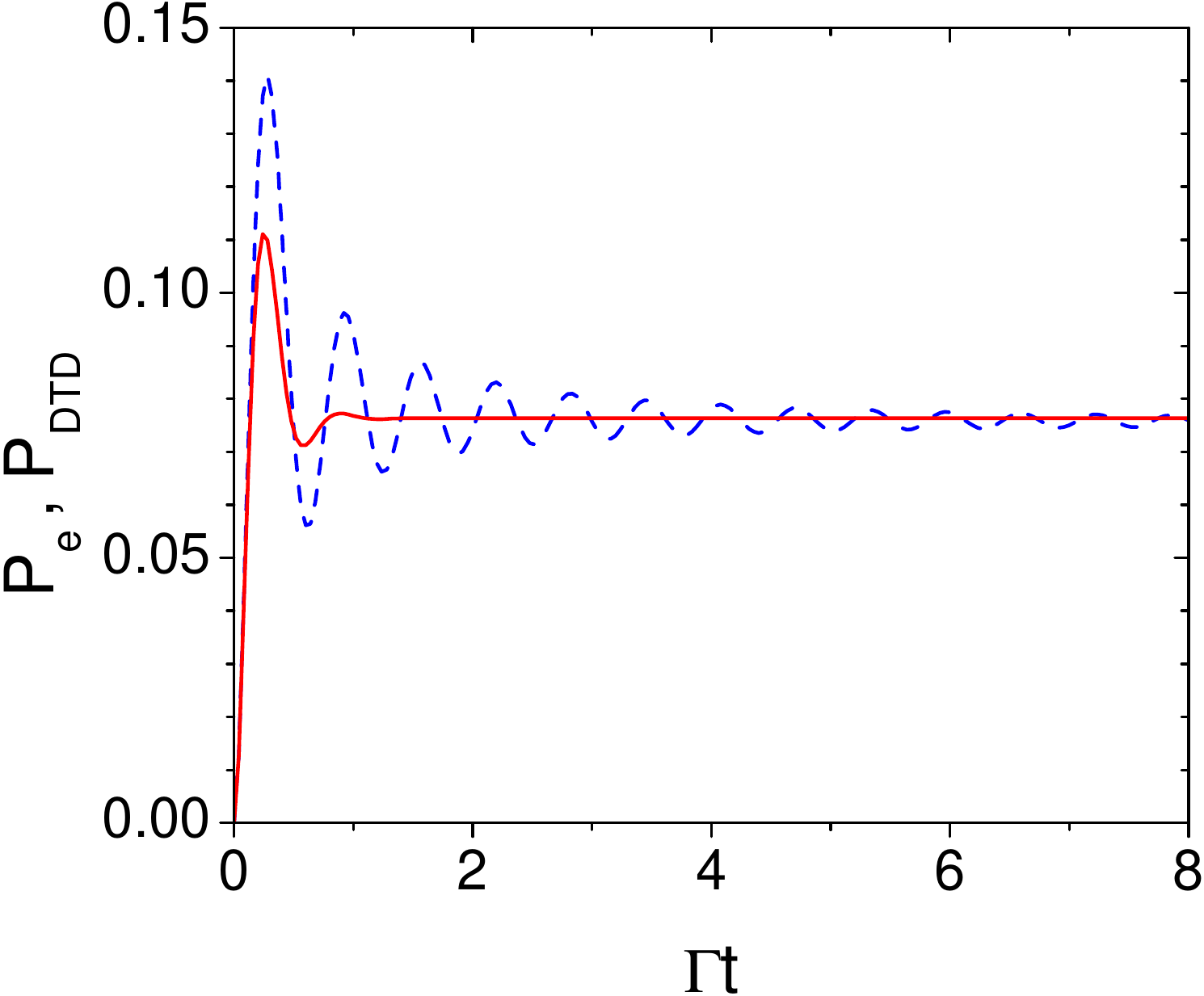}}}
        \caption{(color online) $P_{DTD}(t)=|\beta(t)|^2$ (red solid line) and $P_e=\sum_j|\tilde\beta_j(t)|^2$ (blue dashed line) as a function
        of $\Gamma t$ obtained using the evolution equations (\ref{EqMotion2}) and (\ref{Eqbeta2}), respectively.
        The simulation parameters are $N=4000$, $\sigma=10$ for a gaussian spherical cloud, $\Delta_0=10\, \Gamma$ and $\Omega_0=0.1\,\Gamma$.}
        \label{TDS}
    \end{figure}

\bigskip
\textit{Time evolution of the system --}
 The time evolution of the amplitudes is obtained by inserting
the Hamiltonian (\ref{Ham}) and the ansatz (\ref{EqAnsatz}) into
the Schr\"odinger equation, $\partial_t|\Psi(t)\rangle
=-(i/\hbar)\hat{H}|\Psi(t)\rangle$:
\begin{align}
    \dot{\alpha} =&-\tfrac{i}{2}\Omega_0\sum_{j=1}^N\tilde\beta_j~,\label{EqAmplitude1}\\
    \dot{\tilde\beta}_j =&\,
    i\Delta_0\tilde\beta_j-\tfrac{i}{2}\Omega_0\alpha
        -i\sum_{\mathbf{k}}g_k\gamma_{\mathbf{k}}e^{i(\Delta_0-\Delta_k)t+i(\mathbf{k}-\mathbf{k}_0)\cdot \mathbf{r}_j},\label{EqAmplitude2}\\
    \dot{\gamma}_{\mathbf{k}} =&-ig_ke^{-i(\Delta_0-\Delta_k)t}\sum_{j=1}^N\tilde\beta_j e^{-i(\mathbf{k}-\mathbf{k}_0)\cdot\mathbf{r}_j}.\label{EqAmplitude3}
\end{align}
Integrating Eq.~(\ref{EqAmplitude3}) over time and substituting
$\gamma_{\mathbf{k}}(t)$ in Eq.~(\ref{EqAmplitude2}) we obtain:
\begin{align}\label{Eqbeta}
    \dot{\tilde\beta}_j= i\Delta_0\tilde\beta_j - \tfrac{i}{2}\Omega_0\alpha-\sum_{\mathbf{k}}g_k^2\sum_{m=1}^N
    e^{i(\mathbf{k}-\mathbf{k}_0)\cdot(\mathbf{r}_j-\mathbf{r}_m)}
    \int_0^{t}dt' e^{i(\Delta_0-\Delta_k)(t-t')}\tilde\beta_m(t').
\end{align}
Assuming the Markov approximation (valid for $\tau_N\gg
\sigma_r/c$, where $\sigma_r$ is the size of the atomic cloud and
$\tau_N$ is the cooperative decay time), we can approximate
\begin{align}\label{EqMarkov}
    \int_0^{t}dt' e^{i(\Delta_0-\Delta_k)(t-t')}\tilde\beta_m(t')
    \approx \frac{\pi}{c}\delta(k-k_0)\tilde\beta_m(t).
\end{align}
Then, going to continuous momentum space via
$\sum_{\mathbf{k}}\rightarrow V_{ph}(2\pi)^{-3}\int_0^\infty dk
k^2\int d\mathbf{\Omega_k}$ (where $d\mathbf{\Omega_k}= \sin\theta
d\theta\,d\phi$) and neglecting saturation assuming $\alpha\approx
1$, we obtain
\begin{align}
    \dot{\tilde\beta}_j= i\Delta_0\tilde\beta_j - \tfrac{i}{2}\Omega_0-\tfrac{1}{2}\Gamma
    \sum_{m=1}^N\gamma_{jm}\tilde\beta_m~,\label{Eqbeta2}
\end{align}
where $\Gamma\equiv(V_{ph}/\pi c)k_0^2g_{k_0}^2$ and
\cite{Scully09}
\begin{eqnarray}\label{gjm}
    \gamma_{jm}&=&   \frac{1}{4\pi} \int d\mathbf{\Omega_k}
    e^{i(\mathbf{k}-\mathbf{k}_0)\cdot(\mathbf{r}_j-\mathbf{r}_m)}
    =e^{-i \mathbf{k}_0\cdot (\mathbf{r}_j-\mathbf{r}_m)}
    \frac{\sin(k_0|\mathbf{r}_j-\mathbf{r}_m|)}{k_0|\mathbf{r}_j-\mathbf{r}_m|}
\end{eqnarray}
 Due to the presence of the driving term, the solution will
evolve quickly toward the driven timed Dicke state
\cite{Scully06}, characterized by
\begin{equation}\label{EqTimed}
    \tilde\beta_j(t)=\frac{\beta(t)}{\sqrt{N}}.
\end{equation}
Once inserted the ansatz (\ref{EqTimed}), Eq.~(\ref{Eqbeta2})
yields
\begin{align}
    \dot{\beta}=-\tfrac{i}{2}\sqrt{N}\Omega_0+\left(i\Delta _0-\tfrac{1}{2}\Gamma Ns_N\right)\beta ~,
\label{EqMotion2}
\end{align}
where
\begin{equation}\label{EqsN}
    s_N = \frac{1}{4\pi}\int_0^{2\pi}d\phi\int_0^{\pi}d\theta\sin\theta\left|S_N(k_0,\theta,\phi)\right|^2~.
\end{equation}
and $S_N(\mathbf{k})=
(1/N)\sum_{j=1}^N\exp[-i(\mathbf{k}-\mathbf{k}_0)\cdot\mathbf{r}_j]$
is the structure factor of the atomic cloud. In steady state we
find
\begin{align}\label{EqSteadybeta}
    \beta^{st} & \approx\frac{\sqrt{N}\Omega_0}{2\Delta_0+iN\Gamma s_N}~.
\end{align}
In fig.\ref{TDS} we compare the probability to find atoms in the
driven timed Dicked state (\ref{EqTimed}),
$P_{DTD}(t)=|\beta(t)|^2$ (red continuous line) obtained from Eq.
(\ref{EqMotion2}), and the probability that atoms are excited,
$P_e(t)=\sum_{j=1}^N |\tilde\beta_j(t)|^2$ (blue dashed line)
obtained solving numerically Eq.(\ref{Eqbeta2}), for a spherical
gaussian cloud with $N=4000$ atoms, size $\sigma=k_0\sigma_r=10$
and a pump beam with $\Delta_0=10\Gamma$ and $\Omega_0=0.1\Gamma$.
We observe that the exact state tends toward the driven timed
Dicke state.

\bigskip
\textit{Forces in the Markov approximation.---} The two terms in
the Hamiltonian~(\ref{Ham}) yield two different contributions to
radiation pressure force:
\begin{equation}
    \mathbf{\hat{F}}_{aj}+\mathbf{\hat{F}}_{ej}=-\nabla_{\mathbf{r}_j}\hat{H}~.
\end{equation}
We will be interested in the average absorption force,
$\mathbf{F}_a=\frac{1}{N}\sum_j\langle\mathbf{\hat
F}_{aj}\rangle$, and emission force,
$\mathbf{F}_e=\frac{1}{N}\sum_j\langle\mathbf{\hat
F}_{ej}\rangle$, acting on the center of mass of the whole cloud,
$\mathbf{F}_a+\mathbf{F}_e=m{\mathbf{a}}_{CM}$, where
$\mathbf{a}_{CM}$ is the center-of-mass acceleration and $m$ the
mass of one atom. The first term,
$\mathbf{\hat{F}}_{aj}=\tfrac{i}{2}\hbar\mathbf{k}_0\Omega_0[\hat{\sigma}_je^{i\Delta_0t-i\mathbf{k}_0\cdot\mathbf{r}_j}-\textrm{h.c.}]$
results from the recoil received upon absorption of a photon from
the pump laser and has an expectation value on the timed Dicke
state (\ref{EqTimed}) given by:
\begin{equation}\label{EqAbsorptionforce}
   \mathbf{F}_a=\langle\mathbf{\hat{F}}_{aj}\rangle=-\frac{\hbar\mathbf{k_0}\Omega_0}{\sqrt{N}}\text{Im}\left[\beta(t)\right]~.
\end{equation}
where we assumed again $\alpha\approx 1$. The second contribution,
$\mathbf{\hat{F}}_{ej}=i\sum_{\mathbf{k}}\hbar\mathbf{k}g_k[\hat{\sigma}_j\hat{a}_{\mathbf{k}}^{\dagger}e^{i\Delta_kt
-i\mathbf{k}\cdot\mathbf{r}_j}-\textrm{h.c.}]$, results from the
emission of a photon into any direction $\mathbf{k}$. The
expectation value on the general state (\ref{EqAnsatz}) is:
\begin{equation}\label{EqEmission}
    \langle\mathbf{\hat{F}}_{ej}\rangle=i\sum_{\mathbf{k}}\hbar\mathbf{k}g_k\left[\tilde\beta_j\gamma_{\mathbf{k}}^\ast
    e^{-i(\Delta_0-\Delta_k)t-i(\mathbf{k}-\mathbf{k}_0)
        \cdot\mathbf{r}_j}-\textrm{c.c.}\right].
\end{equation}
Substituting the time integral of $\gamma_{\mathbf{k}}(t)$ from
Eq.~(\ref{EqAmplitude3}) and inserting the timed Dicke state from
Eq.~(\ref{EqTimed}) we obtain the average emission force:
\begin{align}
   \mathbf{F}_e = & -\sum_{\mathbf{k}}\hbar\mathbf{k}g_k^2|S_N(\mathbf{k})|^2
  \left[\beta(t)\int_0^t dt'
   e^{i\left(\omega_k-\omega_0\right)t'}\beta^{\ast}(t-t')+\textrm{c.c.}\right].
\end{align}
\begin{figure}[b]
        \centerline{{\includegraphics[height=3.75cm]{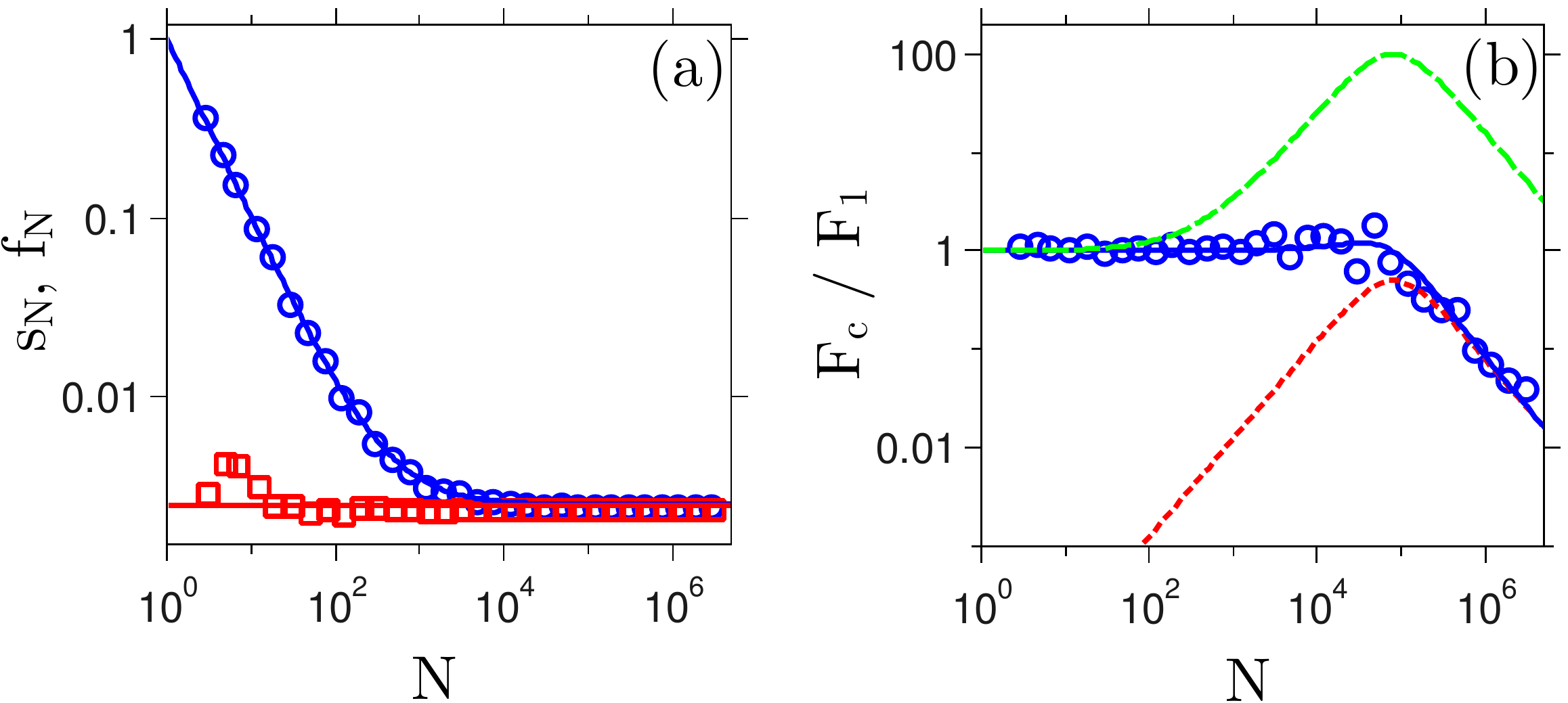}}}
        \caption{(color online)Analytical expressions and numerical
        evaluation for $\sigma=10$ with a configuration average on 10 realizations as a function of atom number.
(a) Results for $s_N$ (blue circles), $f_N$ (red squares) and
analytical expressions $1/N+s_\infty$ (blue line), $f_\infty$ (red
line). (b) Forces acting on a cloud of atoms with
$\Delta_0=-100~\Gamma$: numerical evaluation (blue circles) of the
average cooperative force. The full lines indicate (i) the force
in presence of isotropic scattering, i.e. assuming $f_N=0$ (green
dashed line), (ii) the force for continuous density distributions
without disorder (red dotted line) and (iii) the total force
taking into account cooperative scattering and disorder (blue
line).} \label{Fig2}
    \end{figure}
In the Markov approximation and going to continuous momentum space
we find
\begin{equation}\label{EqEmissionforce}
    \mathbf{F}_e=-\hbar\mathbf{k}_0\Gamma\left|\beta(t)\right|^2 f_N~.
\end{equation}
where
\begin{equation}\label{EqfN}
    f_N = \frac{1}{4\pi}\int_0^{2\pi}
    d\phi\int_0^{\pi}d\theta\sin\theta\cos\theta\left|S_N(k_0,\theta,\phi)\right|^2.
\end{equation}
Finally, using Eq.~(\ref{EqSteadybeta}) in
Eqs.~(\ref{EqAbsorptionforce}) and (\ref{EqEmissionforce}), the
average steady-state radiation force acting on the center of mass
of the atomic cloud is
\begin{equation}\label{Eqforce}
    \mathbf{F}_c \equiv \mathbf{F}_a+\mathbf{F}_e
    = \hbar
    \mathbf{k}_0\Gamma\frac{N\Omega_0^2}{4\Delta_0^2+N^2\Gamma^2s_N^2}~(s_N-f_N).
\end{equation}
The common prefactor can be obtained from the standard low
saturation single-atom radiation force $\mathbf{F}_1 = \hbar
\mathbf{k}_0 \Gamma\Omega_0^2/(4\Delta_0^2+\Gamma^2)$  by
substituting the natural linewidth by the collective linewidth,
$\Gamma\rightarrow N\Gamma s_N$, and the Rabi frequency by the
collective Rabi frequency, $\Omega_0\rightarrow\sqrt{N}\Omega_0$.
Additionally, the cooperative radiation pressure force is weighted
by the difference of structure factors, $s_N-f_N$, where the $s_N$
part corresponds to the cooperative absorption process and the
$f_N$ part to the cooperative emission. For smooth density
distributions $n(\mathbf{r})$, one could compute the structure
functions by replacing the sum with an integral ($s_N\rightarrow
s_\infty$ and $f_N\rightarrow f_\infty$). However, we have shown
\cite{Bienaime10} that in this way we miss the role of the
disorder in the scattering process and the crossover from single
atom radiation force and cooperative scattering. Instead,
estimating the fluctuations of $s_N$ and $f_N$ one finds that (see
fig.\ref{Fig2}a):
\begin{equation}\label{discrete}
s_N \approx \frac{1}{N}+s_\infty,\qquad f_N\approx f_\infty
\end{equation}
Using (\ref{discrete}), the ratio between the cooperative
radiation force (\ref{Eqforce}) and the single atom force is
\begin{equation}\label{ratioF}
\frac{F_c}{F_1}=\frac{4\Delta_0^2+\Gamma^2}{4\Delta_0^2+\Gamma^2(1+Ns_\infty)^2}
\left[1+N(s_\infty-f_\infty)\right]
\end{equation}
Assuming a smooth Gaussian density distribution with ellipsoidal
shape,
$n_0\exp[-(x^2+y^2)/2\sigma_r^2-z^2/2\sigma_z^2]$, the
structure factor is
$S_\infty(k_0,\theta,\phi)=\exp\{-\sigma^2[\sin^2\theta+\eta^2(\cos\theta-1)^2]/2\}$,
where $\sigma=k_0\sigma_r$ and $\eta=\sigma_z/\sigma_r$ is the
aspect ratio. For elongated clouds, $\eta\ge 1$,
\begin{align}
    s_\infty^{(\eta)} & = \frac{\sqrt{\pi}e^{\frac{\sigma^2}{\eta^2-1}}}{4\sigma\sqrt{\eta^2-1}}\left\{\text{erf}\left[\frac{\sigma(2\eta^2-1)}
        {\sqrt{\eta^2-1}}\right]-\text{erf}\left[\frac{\sigma}{\sqrt{\eta^2-1}}\right]\right\}~,\nonumber\\
    f_\infty^{(\eta)} & = \frac{1}{\eta^2-1}\left[\eta^2 s_\infty^{(\eta)}-\frac{1}{4\sigma^2}(1-e^{-4\eta^2\sigma^2})\right].
\end{align}
For spherical clouds ($\eta=1$) and for $\sigma\gg 1$ one finds
$s_\infty \approx 1/4\sigma^{2}$ and $s_\infty -f_\infty\approx
1/8\sigma^{4}$. For $\sigma,\eta\gg1$, $s_\infty^{(\eta)}$ can be
approximated by $s_\infty^{(\eta)}\simeq
s_\infty\sqrt{\pi}Fe^{F^2} \left[1-\text{erf}(F)\right]$, where
$F\equiv\sigma/\eta=k\sigma_r^2/\sigma_z$ is the Fresnel number.
For large Fresnel numbers $s_\infty^{(\eta)}\rightarrow s_\infty$.

As illustrated by fig.\ref{Fig2}b, the single-atom force is
recovered in the limit of $N s_\infty\sim N/4\sigma^2\ll 1$ i.e.
for small optical thickness $b_0\approx 3N/\sigma^2$. On the
contrary, for large $b_0$ the microscopic inhomogeneities can be
neglected and cooperativity strongly modifies the radiation force.
In particular, for small volumes the emission is isotropic and
$f_\infty\approx 0$, whereas for large volumes the recoil at the
emission compensates the recoil at absorption, $f_\infty\approx
s_\infty$, which results in mainly forward emission.

\section{Experimental Results}

The analytical results presented in the previous section should apply to a large variety
of clouds of cold atoms, including thermal cold atoms as well as degenerate
quantum gases (as long as atom-atom interactions can be neglected).
In this section we report first observations of cooperative scattering
in different regimes : thermal clouds in a dipole trap or released from
a dark MOT and for a Bose-Einstein condensate realized from a magnetic
trap.

\subsection{Thermal Cold Atoms in a Dipole Trap}
\label{Nice2008}

\begin{figure}[h]
        \centerline{\scalebox{0.25}{\includegraphics[angle=90]{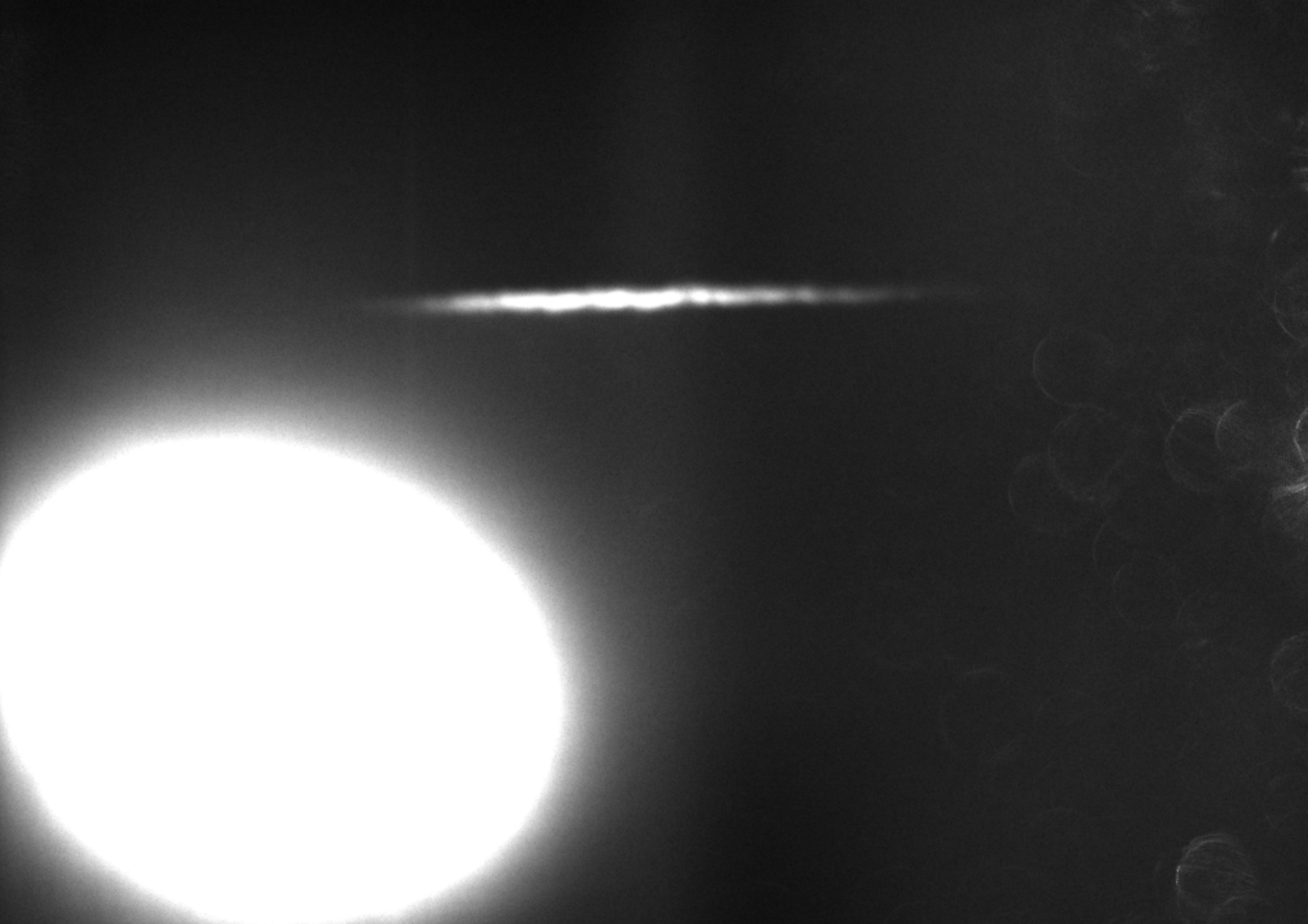}}}
        \caption{Fluorescence image of the cold atoms. The large cloud corresponds to the free falling atoms and the narrow cigar shaped cloud to atoms trapped in the dipole laser.}
        \label{FigNiceMOTDipole}
    \end{figure}

A first series of experiments have been performed in Nice in 2008, where we were looking for possible signatures related to collective atomic recoil lasing (CARL), with
spontaneous atomic bunching,  when a large cloud of atoms is exposed to off resonant detuned light. In contrast to previous experiments \cite{Kruse03}, this setup does not use a high finesse cavity, but the larger atom number we are able to trap \cite{Labeyrie06} might allow to compensate for the absence of the cavity. The results of these studies did not shown any evidence of CARL, but provided the first signatures of cooperative scattering.
The experiment has been performed using a vapor loaded magneto-optical trap (MOT) of $^{85}$Rb atoms. The cooling laser has been derived from an DFB master laser amplified by a tapered amplifier (TA), whereas the repumper laser has been a simple DFB laser. After a dark MOT period of 35ms, we load the atoms into a red detuned single beam dipole trap, formed by another DFB laser amplified by a TA and focused to a beam waist of $\approx 200\mu m$. The action of this dipole trap laser is twofold: on one side it holds atoms against gravity due to the dipole forces and on the other side, as the atom-laser detuning is not very large, the residual radiation pushes the atoms along the dipole trap. As one can see in Fig.~\ref{FigNiceMOTDipole} most of the atoms of the dark MOT are not loaded into the dipole trap and fall under the action of gravity. A small fraction of the atoms are however trapped in the dipole laser beam and are pushed along the axes of propagation of the laser. This observation is not surprising, given the moderate detuning (at least for dipole traps) we have used, with values ranging from $\Delta_0=-50GHz$ to $\Delta_0=-200GHz$ (detuning $\Delta_0$ given in respect to the $F=3\rightarrow F'=4$ transition of the D2 line of $^{85}$Rb).
This regime of dipole traps where radiation pressure cannot be neglected is not commonly studied, as spontaneous scattering of photons is usually not desired. The experimental protocol using the dipole trap to both hold the atoms against gravity and push them with off resonant radiation pressure made it difficult to independently change the parameters for the dipole trap (and thus the size and shape of the atomic cloud) and for the radiation pressure effects. This experiment however provided our first signatures of cooperative scattering which have been studied later on in a more quantitative way. We thus show in this section our first qualitative results, which could not be compared in a quantitative way to a theoretical model. As this experimental protocol required a finite interaction time in the dipole trap in order to allow separation of the untrapped and trapped atoms, the shape of the clould of atoms changes during this interaction time. This further complicates a reliable comparison to theoretical model.

    \begin{figure}[h]
        \centerline{\scalebox{0.3}{\includegraphics{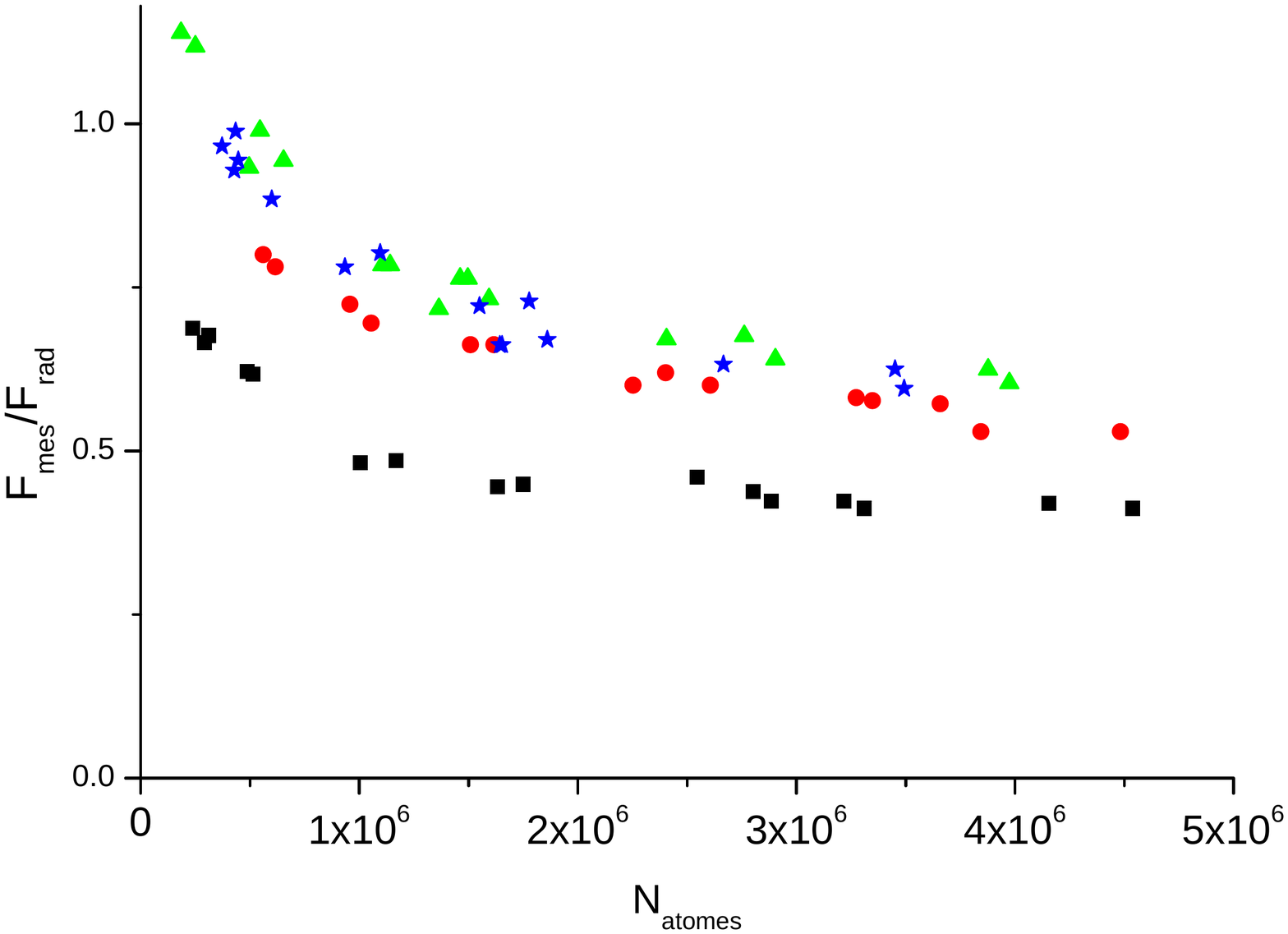}}}
        \caption{(color online) Average radiation force on the cloud of the atoms. For increasing atom number, the cloud is less displaced by the dipole laser. The interaction time for this experiment is $50ms$, the laser beam power is $100mW$ for $\Delta_0=-76GHz$ (black square), $\Delta_0=-87GHz$ (red circles), $\Delta_0=-90GHz$ (green triangles) and $\Delta_0=-108GHz$ (blue stars).}
        \label{FigNiceFN}
    \end{figure}

    \begin{figure}[h]
        \centerline{\scalebox{0.3}{\includegraphics{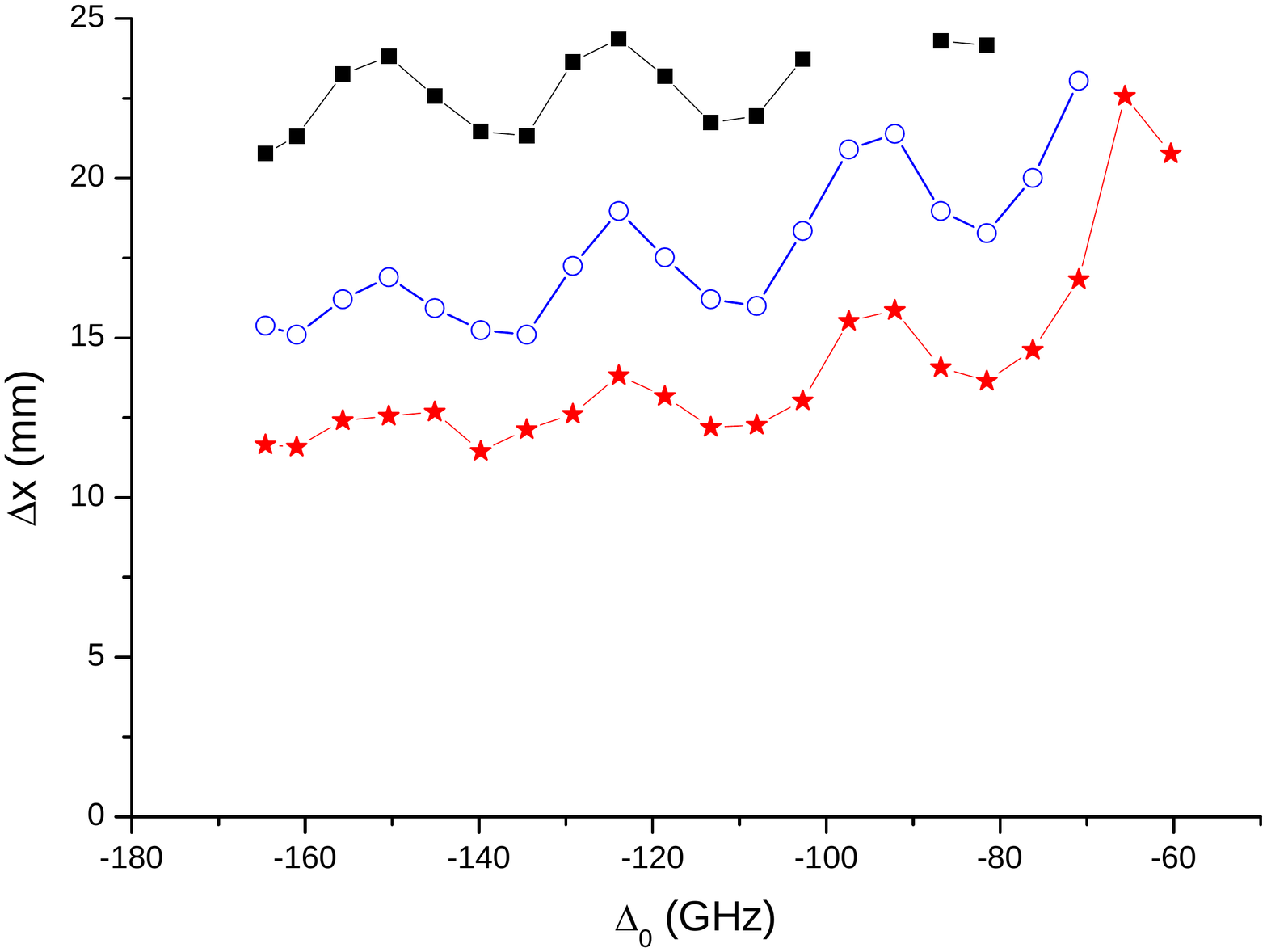}}}
        \caption{(color online) Displacement of the center of mass of the atomic cloud as a function of detuning for different interaction times: $50 ms$ (red stars), $60 ms$ (open blue circles), $70 ms$ (black squares) obtained for a laser power of $P=100mW$.}
        \label{FigNiceOsc}
    \end{figure}

In Fig.~\ref{FigNiceFN} we plot the average force acting on the center of mass of the cloud of atoms kept in the dipole trap. This force can be extracted from the spatial displacement, as the interaction  time is well known. The normalization of the force to the single atom force is roughly estimated from the measured values of laser power and detuning. The most striking point to notice in  Fig.~\ref{FigNiceFN} is the clear reduction of the average radiation force with increasing atom number. This effect has been subsequently been studied in a quantitative way \cite{Bienaime10}.

During the course of these experiments, we have observed some intriguing features which have not yet been studied in a quantitative way. For instance we found a systematic oscillation of the average radiation pressure force, as illustrated in Fig.~\ref{FigNiceOsc}. This feature clearly merits further experimental investigation, as if it is not due to an experimental artefact,
it might be related to effects beyond the Markhov approximation, where ringing of superradiant time decay is predicted \cite{Svidzinsky08}.

\subsection{Thermal Cold Atoms released from a dark MOT}
\label{Nice2009}

\begin{figure}[ht]
        \centerline{\scalebox{0.5}{\includegraphics{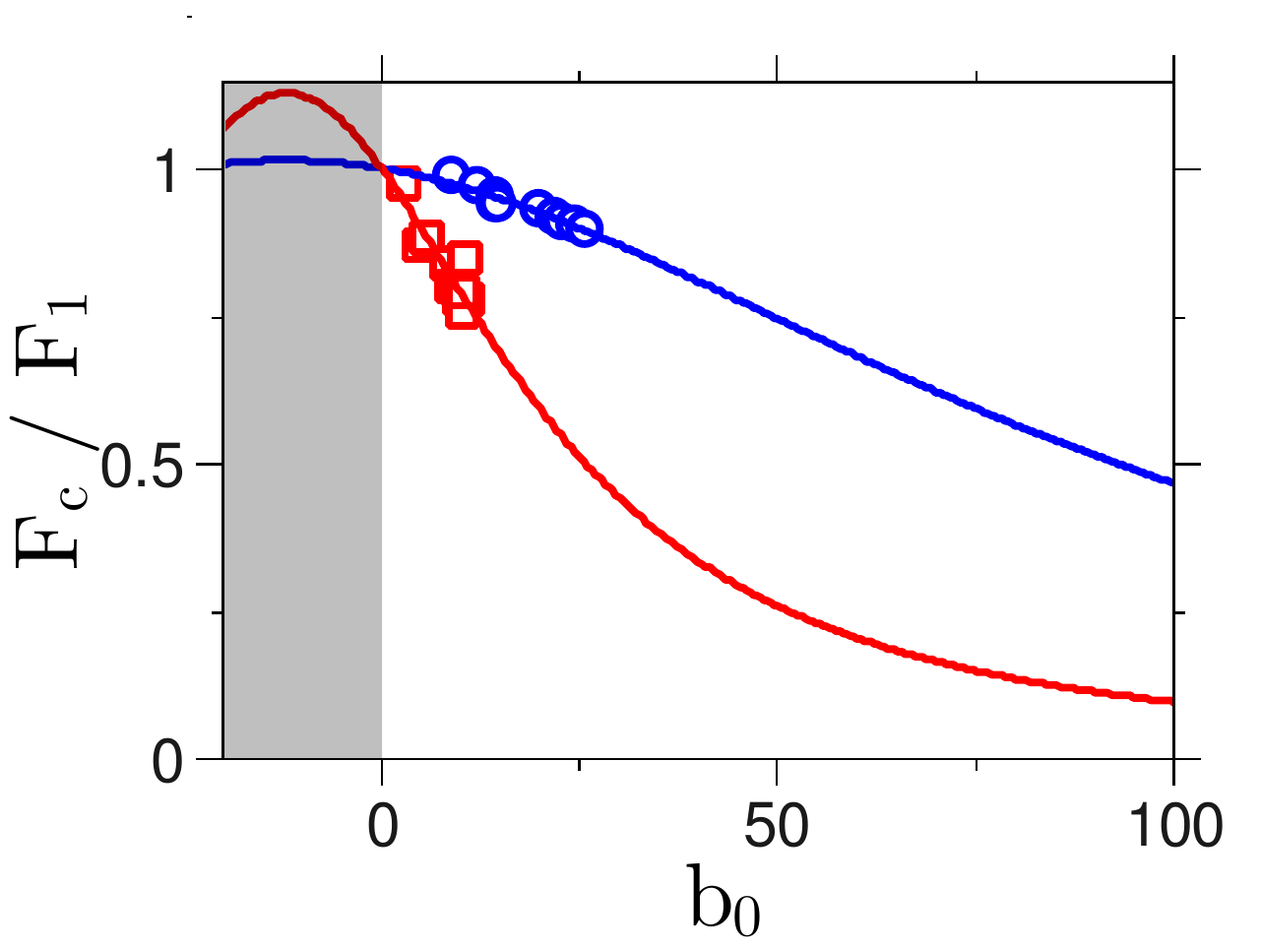}}}
        \caption{(color online) Experimental data and fits using the cooperative radiation force $F_c$ (normalized to the single atom radiation force $F_1$) in presence of disorder for $\Delta_0=-1.9 \, \Gamma$ (red squares) and $\Delta_0=-4.2 \, \Gamma$ (blue circles). The shadowed area corresponds to the non-physical region $b_0 < 0$.}
        \label{FigNice2009}
    \end{figure}

Following the first series of experiments and the development of our theoretical model presented above, we performed experiments which allowed for a quantitative comparison. The main results of these experiments have been presented in \cite{Bienaime10} and we thus only show the main result here. These experiments have been done using the same vapor cell as in the previous experiments. These new results have however used the $^{87}$Rb isotope. The experimental protocol used to allow for a quantitative measurement of cooperative scattering did not use atoms in a dipole trap, but atoms released from a dark MOT. Adapting the number of atoms interacting with the pushing laser by controlled repumping from the $F=1$ to the $F=2$ hyperfine level in the ground state, we changed the optical thickness of the cloud without changing the size of the cloud. This allowed for a quantitative measurement of the radiation force as illustrated in Fig.~\ref{FigNice2009}. We take the good agreement between experiments and theory as a proof of the relevance of cooperative scattering in clouds of cold atoms.

\subsection{Measurements in a magnetic trap}

The experiment described above present the first clean signatures of cooperative scattering of single photons along the lines of the model outlined in Refs.~\cite{Courteille09,Scully06,Svidzinsky08}. At the large pump laser detunings and the large cloud volumes used, the light was strongly scattered into forward direction.
In the case of small volumes, the absorbed light can be reemitted in directions other than the forward one. Cooperativity then also leaves its imprint in the geometry of the radiation pattern.

Particularly small and dense samples can be made by cooling the atomic cloud to ultralow temperatures. Ultimately, their size is limited by the repulsive interatomic interaction making it difficult to reach sample sizes on the order of $\sigma\simeq1$. Another advantage of samples with temperatures below the recoil limit is, that the interaction with a light field leaves detectable traces in their momentum distribution even when on average every atom scatters much less than a single photon.

\begin{figure}[ht]
		 \centerline{\includegraphics[width=4cm]{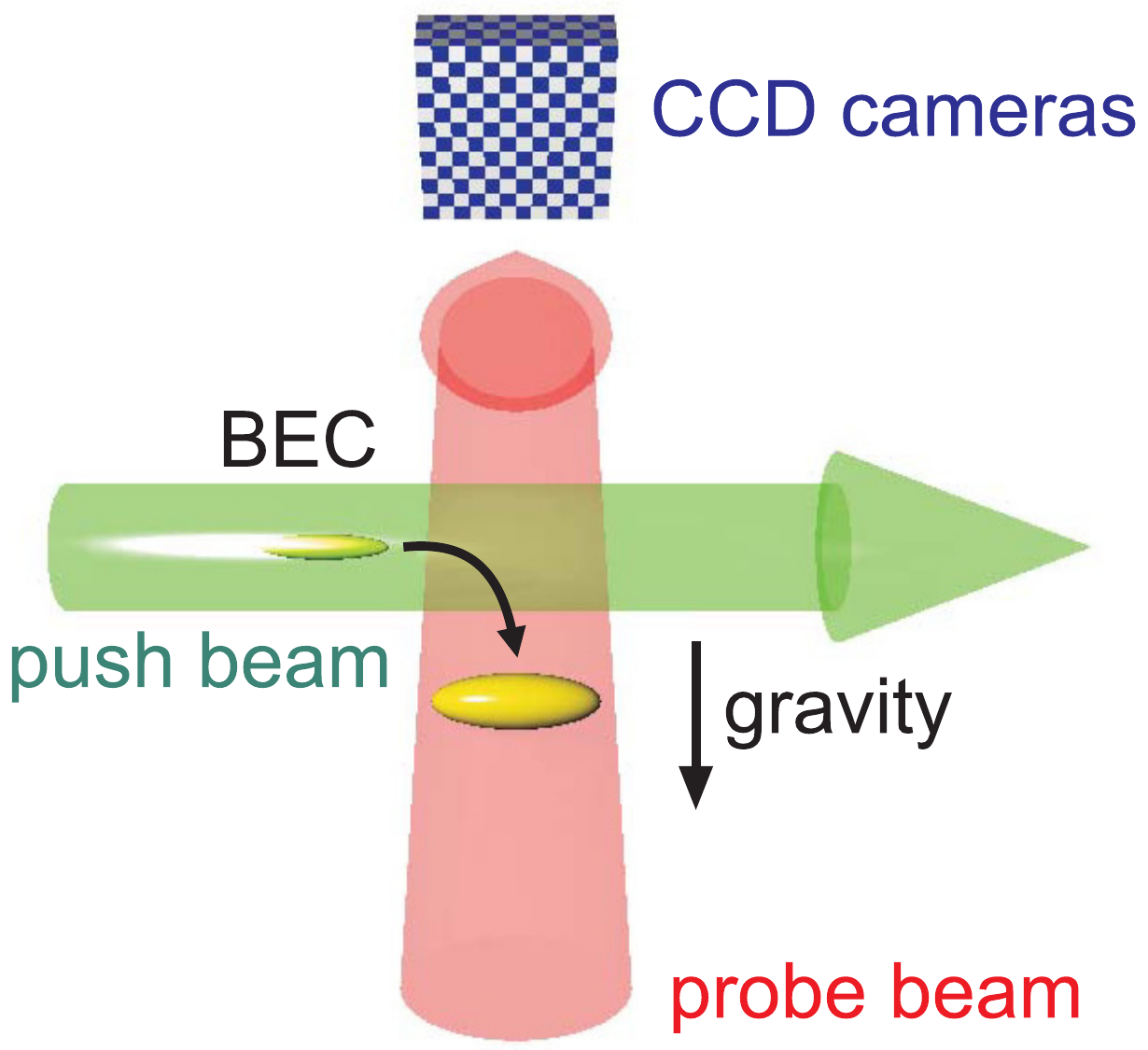}}
		\caption{(color online) Scheme of the experiment.}
		\label{FigCS_Bec_Setup}
	\end{figure}

In an experiment performed at the university of T\"ubingen we prepare a $^{87}$Rb cloud in a magnetic trap and cool it down by forced evaporation to quantum degeneracy. The trap is ellipsoidal with frequencies $\omega_z/2\pi=25~$Hz and $\omega_r/2\pi=160~$Hz. We now apply in axial direction a light pulse with a power of $P_0=130~\mu$W and a diameter of $w_0=295~\mu$m for a duration of $\tau_0=10...500~\mu$s detuned from the D2 line by $\Delta_0/=\pm0.5...\pm4~$GHz. Immediately after the pulse (within $100~\mu$s) the trap is switched off. The cloud falls in free expansion for $t_{tof}=20~$ms before we apply an imaging pulse in transversal (radial) direction (see Fig.~\ref{FigCS_Bec_Setup}).

\begin{figure}[ht]
		 \centerline{\includegraphics[width=10cm]{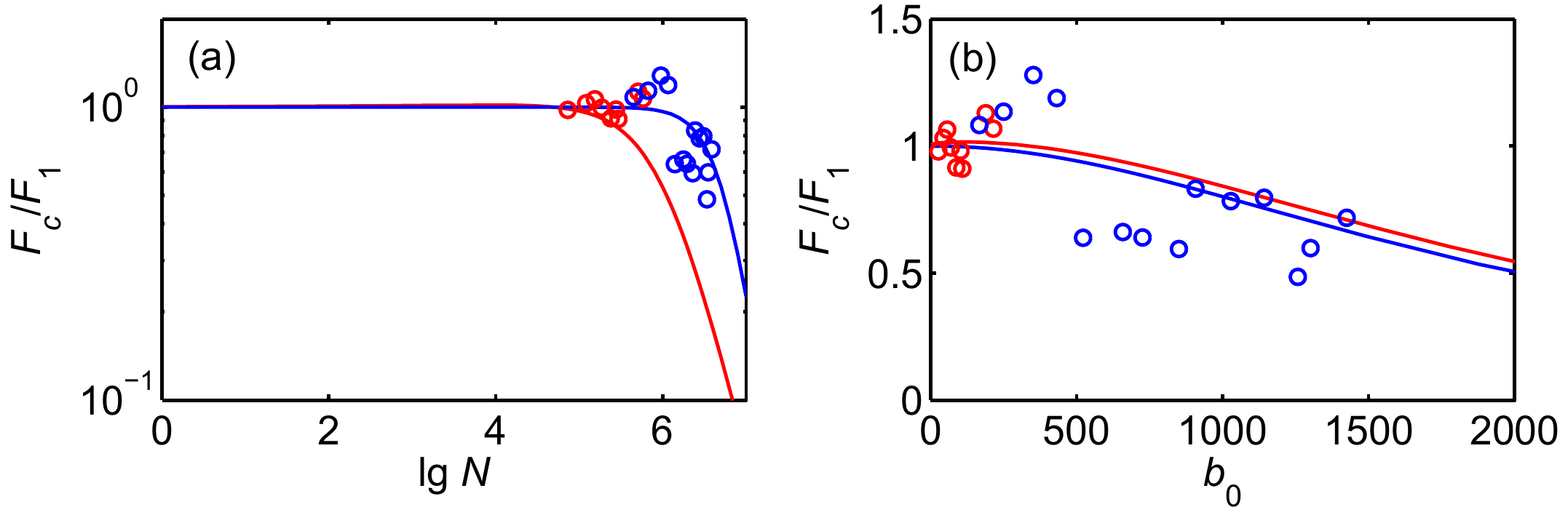}}
		\caption{(color online) Blue lines and symbols correspond to $T=1~\mu$K cold thermal clouds (size and
			shape independent on $N$), red ones to condensates in the Thomas-Fermi limit (size and shape depend on $N$).
			(a) Calculated (lines) and measured (symbols)
			$N$-dependence of the radiation pressure ratio. (b) Same as (a), but in linear scales and as a function of resonant optical density.
			The experimental parameters were $\Delta_0=500~$MHz, $\sigma_+$-pol, $\tau_0=20~\mu$s,
			$\tau_{tof}=20~$ms. No parameter were adjusted. Every data point is an average over several measurements.}
		\label{FigMessung091217}
	\end{figure}

We either use thermal clouds of $N=10^5...4\cdot10^6$ atoms at a temperature of $T\simeq~1~\mu$K or Bose-Einstein condensates of $N=10^4...6\cdot10^5$ atoms. In the case of thermal clouds the aspect ratio is $\eta=5.6$ and the size is $\sigma=90$, independent on $N$ (see Fig.~\ref{FigMessung091217}). In the case of a condensate the interatomic interaction gives rise to a chemical potential of up to $\mu/h=10~$kHz and a transverse radius in the Thomas-Fermi limit of up to $\sigma=60$. Also the aspect ratio depends on the atom number, because the mean field presses the condensate into the weakly confining dimension. Optical densities of up to $b_0=2000$ are reached in axial direction.


\bigskip

\begin{figure}[h]
		 \centerline{\includegraphics[width=4cm]{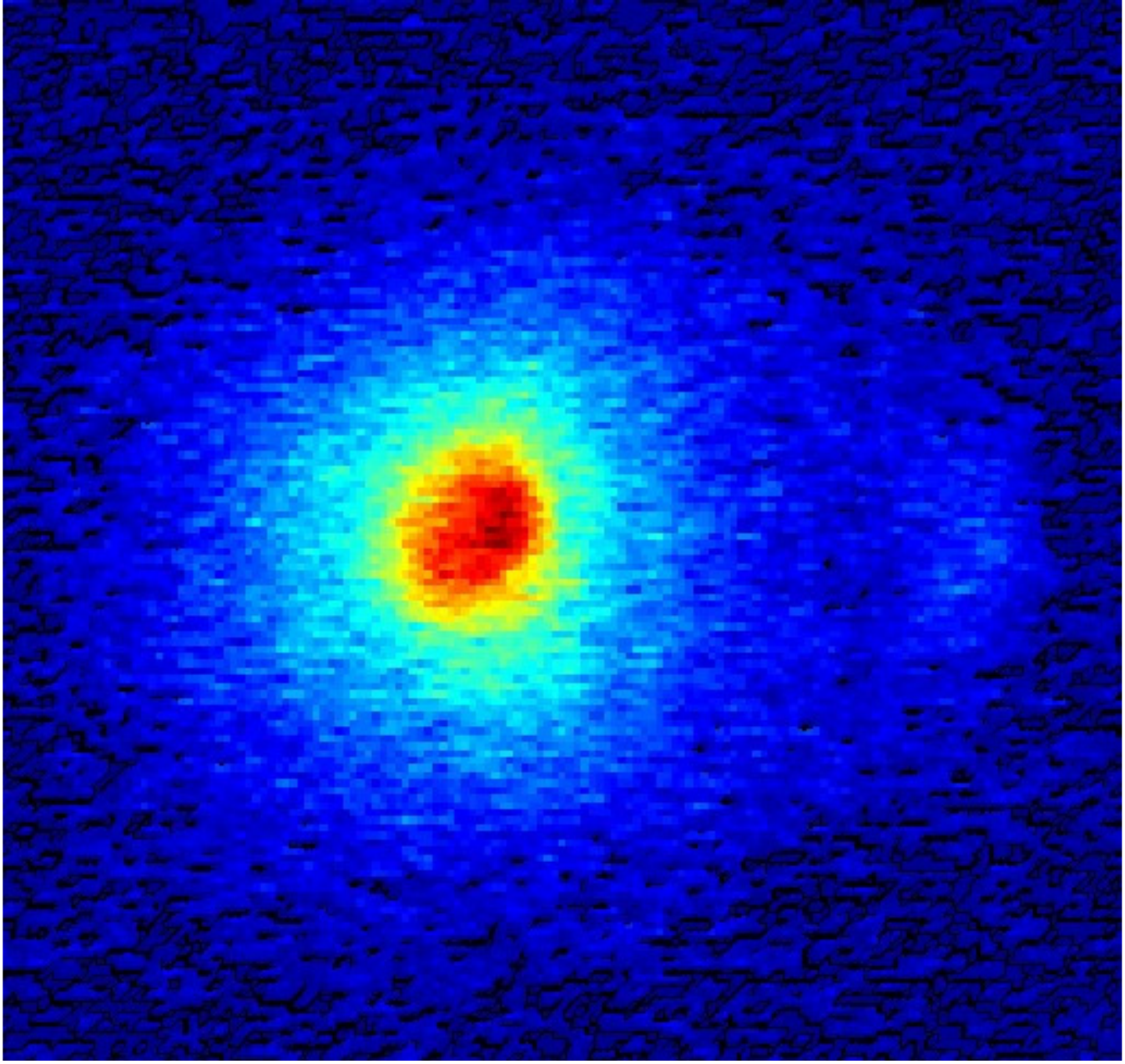}}
		\caption{ (color online) Absorption images of atomic clouds released from the magnetic trap.}
		\label{FigMessungHalo}
	\end{figure}

The effective radiation pressure is extracted from time-of-flight absorption images such as the one shown in Fig.~\ref{FigMessungHalo}.
The first moment of the momentum distribution $\Delta p=m\Delta z_{cm}/t_{tof}$ is a measure for the collective radiation force.
The tendency of the radiation force to decrease with increasing atom number is clearly visible and will be subject to future investigations.

\section{Conclusion}

Cooperative effects in scattering of light by large clouds of cold atoms present phenomena which can be described by using driven timed Dicke states. These states present a convenient quantum approach even though the features exploited in the present paper do not go beyond what can be expected from a classical treatment. We have given detailed results on our theoretical model, with an analytical expression for the modified radiation force when large clouds of atoms are used. Experimental confirmation of the modification of the radiation force has been observed in two different laboratories using different experimental configurations, ranging from cold atoms held in an optical dipole trap, atoms released from a dark MOT to atoms in a BEC setup with a magnetic trap. These experiments illustrate the wide range of situations where such cooperative scattering processes need to be considered.
Important future extensions of this work arising from this cooperative processes would include any possible 'quantum' feature which could not be described with classical models. Fluctuations and higher order correlations seem appropriate first signatures to study in this respect.

\section{Acknowledgements}

This work has been supported by ANR (projects ANR-06-BLAN-0096 and ANR-09-JCJC-009401), DFG (Contract No. Co~229/3-1), DAAD and INTERCAN.

\label{lastpage}

\end{document}